\documentclass{article}

\usepackage{graphicx}
\usepackage{comment}

 \usepackage[main, final]{iaseai26}

\usepackage[utf8]{inputenc} %
\usepackage[T1]{fontenc}    %
\usepackage{hyperref}       %
\usepackage{url}            %
\usepackage{booktabs}       %
\usepackage{amsfonts}       %
\usepackage{nicefrac}       %
\usepackage{microtype}      %
\usepackage{xcolor}         %
\usepackage{graphicx}
\usepackage{natbib}

\usepackage{natbib}
\setcitestyle{numbers,square}

\setcitestyle{numbers,square}

\usepackage{enumitem}
\newlist{inlinenums}{enumerate*}{1}
\setlist[inlinenums,1]{label=\textbf{(\arabic*)}, itemjoin={{}}}

\newtoggle{comments}  %
\settoggle{comments}{true}
\iftoggle{comments}{

}

\title{AI Safety Evaluations Need To Consider\\Cascading Effects}

\author{%
  Anna Neumann \\
  Research Center Trust, UA Ruhr \\
  University of Duisburg-Essen, Germany \\
  \texttt{anna.neumann1@uni-due.de} \\
  \And
  Jatinder Singh \\
  Research Center Trust, UA Ruhr \\
  University of Duisburg-Essen, Germany \&  \\
  University of Cambridge, United Kingdom \\
  \texttt{jatinder.singh@uni-due.de} \\
}

\begin{document}

\maketitle

\begin{abstract}

AI systems comprise a range of interactions across the technical and organisational components of a range of actors. 
 These components work together to provide the systems' functionality. This socio-technical assemblage is increasingly described as an \textit{algorithmic supply chain}. 

Given their role in supporting a wide range of systems, foundation models (FMs) are increasingly a key part of many algorithmic supply chains.
In practice, various technical and non-technical \textit{components}
work to mediate, adapt, and augment the behaviour of
models, such as FMs,
both in general, and for their use in specific application contexts.
However, many AI safety evaluations tend to focus on capabilities of FMs themselves and/or assess these components independently,  at certain levels of abstraction,

\textit{with less consideration on how these components
could interact, influence, reinforce or counteract each other.}

In this paper, we \textbf{introduce the \textit{cascade} as a concept for supporting more holistic AI evaluations}. 
The \textit{cascade} captures
how the interactions between socio-technical components along the AI supply chain can compound and produce cumulative effects with downstream consequences.
Specifically we \textit{(i)} identify gaps in current AI auditing approaches, using LLMs as our case study; \textit{(ii)} demonstrate how cascade problems manifest in deployed AI systems through a characterisation of cascades through different viewpoints 
({component} and stakeholder), 
 and \textit{(iii)} propose research directions for assessing the cascade specifically as an object of analysis.
As these cascades can significantly impact transparency, accountability, security, and safety, \textbf{we advocate for a paradigm shift in AI system auditing towards systems-oriented audits that incorporate cascading effects}, to \textit{complement} model centric evaluations. 
\end{abstract}

\section{Introduction} \label{sec:introduction}

Modern AI systems %
entail complex supply chains~\cite{cobbe_understanding_2023, doi:10.1177/20539517231177620, balayn_stem_2024}: interconnected sets of technical components%
, services%
, and actors working to deliver functionality, from model training to deployment. %

\begin{figure}[ht]
    \centering
    \includegraphics[width=0.9\linewidth]{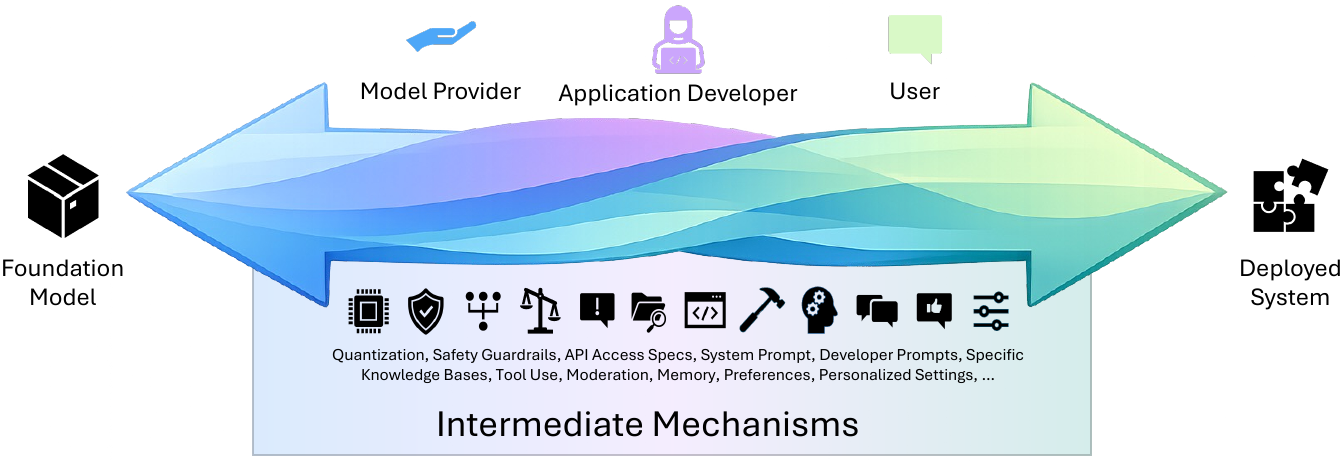}
    \caption{A conceptual illustration an AI supply chain depicting the interactions between a FM and a {user-facing} application (`deployed system')
    between which we see multiple components influenced by different stakeholders: here the  %
     model provider, application developer, and user.}
    \label{fig:intro}
    \vspace{-0.5em}
\end{figure}

To illustrate, consider an LLM-based customer service chatbot: the FM is provided by an organization, hosted by another, integrated with safety guardrails and specific knowledge by a client developer, and personalized through user interaction. 
Each actor---model providers, application developers, end-users%
---controls different \textit{components}. These components can interact in different ways, producing \textit{`flow-on effects'}, where outputs, data flow changes, applied filters, etc, can feed into and otherwise affect other components in terms of how they operate or the data they process (see \autoref{fig:intro}).

These dynamics are becoming increasingly prevalent, which reflects the service-based nature of technologies.
\textit{`AI as a Service'} (AIaaS)  %
is where organisations offer machine learning capabilities as `ready-to-use' components %
to customers~\cite{lewicki_out_2023}. Here, application developers submit inputs and receive outputs (predictions, classifications, generations) that work to drive their applications, %
without necessarily having insight into or oversight over model internal workings and specifics~\cite{cobbe_aiaas_2021}. 

As FMs play an increasingly prominent role as shared components across many AI services and applications~\cite{yang_foundation_2023,  dominguez_hernandez_mapping_2024}, they represent core AIaaS offerings that can provide broad capabilities across multiple domains \citep{zhou_comprehensive_2024, bommasani_opportunities_2022, suresh_participation_2024}. The adaptation of FMs %
to specific applications is mediated through a variety of components in their post-training deployment by \textit{various stakeholders}. %
For generative AI systems, such components can include efficient inference (e.g., quantization, pruning) \cite{kirsten2024the}, prompting strategies \cite{10.1145/3658644.3670388, Ranade2025-qw, 10.1145/3544548.3581388}, %
personalization \cite{noauthor_memory_2024, salemi2024lamplargelanguagemodels, 10.1145/3696410.3714842}, or agentic AI functionalities~\cite{chan_mle-bench_2024, liu_agentbench_2023, noauthor_computer-using_nodate, noauthor_building_nodate}. %

Importantly, each of these components can be potentially managed by different stakeholders. However, the visibility of each actor is typically limited, as one can generally only see (and control) that which they directly manage, but not those components managed by others that may operate up\slash downstream from their own. 

The above describes how a range of technologies and stakeholders interact to deliver functionality as part of AI supply chains: systems affect each other in various upstream or downstream ways with FMs as a core component, without each party necessarily having the visibility or control over the functionality or behaviours of others, let alone a holistic view of the entire chain.

\subsection{Contributions}

Many AI safety\slash security\slash ethical evaluations tend to focus on FMs.
In this paper we highlight how current auditing practices predominately concern evaluating FMs in isolation, or assessing deployed applications (mostly) as black boxes. %
While these capture the dynamics pertaining to outputs, the structure of \textit{how} the components interacted or \textit{why} particular effects emerged is largely overlooked.

We argue the \textbf{importance for audits that explore the \textit{cascade}} of interactions that can occur across technical\slash socio-technical components, stakeholders, and other aspects that drive an AI system. Whether cascades are accounted for {as an object of analysis in} an audit process can directly shape the types of safety, responsibility, and accountability aims that can be realised through the audits, including where an issue occurred, who or what contributed to its occurrence, who should fix it, who is responsible for remediation and recourse, and so on. %

Specifically,  we introduce and analyse cascades along two overlapping modalities, capturing
distinct but interconnected dynamics in socio-technical AI systems:
\textit{(i)} the \textit{component cascade}, where behaviour compounds through connected components, and \textit{(ii)} the \textit{stakeholder cascade}, where different actors in the supply chain influence different aspects of system behaviour. While our argument applies broadly to a range of socio-technical systems in general, given their  composite nature, our focus here is on FMs -- particularly Large Language Models (LLMs). We use LLMs as our case study since LLMs are widely deployed, prominent in the public and governance discourse, and are often key targets of auditors. %

From this analysis, we conclude by calling for \textbf{AI audits that consider the components and interactions %
that form algorithmic supply chains (i.e. \textit{the cascade}).} This is timely and urgent, given that the current auditing landscape is inherently fragmented by %
mostly considering FMs (and their capabilities), or deployed systems as a whole. As such, the components and outcomes in between  are often overlooked, which brings implications for the safety\slash security aspects they surface.
We further highlight concepts generally not considered in supply chain discourse, and outline initial research directions for cascade-aware auditing practices.

\subsection{Definitions} \label{sec:definitions}
{We use the term \textit{components} to refer to the socio-technical elements in AI supply chains that work to shape system behaviour. %
In a FM context, components include mechanisms that can modify, constrain, or augment how models process information and generate outputs, such as system prompts, safety filters, retrieval systems, fine-tuning tools, and user preferences.}
Importantly, what constitutes a component depends on the level of visibility and abstraction being examined. What appears as a single component may actually consist of multiple interacting components %
depending on one's level of access and abstraction; e.g. from an external auditor's perspective, a content moderation API might appear as a single component. However, with internal access, this same system looks like multiple components: the FM, post-processing filters, review queues, and customization layers. 

An \textit{application} refers to a system that delivers end-to-end functionality to users, composed of multiple interacting components, e.g., a customer service AI system or medical diagnosis assistant.
Importantly, what is considered an application depends on the level of abstraction being considered.
That is, \textit{a component from one stakeholder's perspective, might be considered an application by another}: again, that content-moderation API might be an application from the external auditor's  perspective, a single component from an internal developer's perspective, and a series of components from the API provider's perspective. %

{
Each component can give rise to particular \textit{effects} on inputs, processing, or outputs. %
These effects can include transformations, instructions, sensor readings, data queries, or guardrail settings, to name but a few. %
In this paper, when referring to a \textit{transformation}, we refer to  one type of effect that a component can produce; and given our focus on FMs, specifically those that impact data or model behaviour. This could be, for example, the changing of parameters, data, or model instructions. We note that we use transformations as one example of possible effects (\S\ref{sec:cascaudtech}) to streamline and aid our discussions. In practice, many other types of effects (like access controls, API rate limits, {caching policies, routing,} %
and so on), have to be considered.

\vspace{-0.5em}
\section{Background: Seeing (through) AI Supply Chains} \label{sec:aisupplychains}

Recent research has focused on 
AI supply chains from multiple angles, i.e. mapping ecosystems of models and their variants 
\cite{Bommasani_Soylu_Liao_Creel_Liang_2024, laufer2025anatomymachinelearningecosystem}, mapping stakeholders and their relationships \cite{10.1145/3715275.3732017, balayn2025unpacking}, and characterizing AI supply chains (or `value chains' \cite{attardfrost2024ethicsaivaluechains}) as fundamentally non-modular %
\cite{cen2023ai}. Non-modularity means here that the combination of components can lead to emergent effects, just through the act of combining functions (see below). 

In practice, algorithmic supply chains function as interconnected \textit{systems-of-systems}. This impacts comprehensive audits, as visibility over data flows is lost when crossing technical or organisational boundaries, and components are managed by different entities and may behave in ways that make it difficult to determine how particular inputs lead to particular outputs~\cite{singh_decisionprovenance}. A lack of transparency %
has already been noted for multiple parts of the FM ecosystem~\cite{bommasani2023foundation}. Further, there is evidence of performance degradation on sequential chains of small AI models \cite{hopkins2025aisupplychainsemerging}, and work examining how accountability becomes dislocated across these chains \cite{cobbe_understanding_2023, doi:10.1177/20539517231177620, brown2023allocating}. 
Moreover, arguments have already been made around AI safety not being a `model property'~\cite{normaltechSafetyModel}, and that context around the model shapes behaviour and therefore needs to be considered~\cite{Singh2016ResponsibilityM}. This all motivates our consideration of \textit{components} and their interactions as part of a broader audit ecosystem}.
Three factors of AI supply chains are particularly worth mentioning for their \textit{impact on audit feasibility}:

\textbf{Non-modularity.} AI supply chains behave much like a `blended soup' (borrowing the metaphor from \cite{cen2023ai}). In some instances, supply chains can  be compartmentalised into %
parts more like a salad, where the specific ingredient can be easily taken out of the bowl. In AI supply chains, there is often a degree of non-modularity, which means that the `ingredients' (components in the AI supply chain) are blended.
For example, there can be interactions between components that 
 can combine in non-linear ways and influence each other with or through loops, conditional integrations, or repeated calls to specific APIs~\cite{singh_decisionprovenance}. 
As such, there will often be situations where it will not be clear which %
parts contributed if something `tastes off', and also no reliable way to identify a specific ingredient. 
Importantly, it might even be the {particular combination of components and their interaction} %
that makes something `taste off'.

\textbf{Opacity.} Algorithmic supply chains \cite{doi:10.1177/20539517231177620, cobbe_understanding_2023, cen2023ai, brown2023allocating} tend to involve some inherent opacity as %
each stakeholder typically only has limited view of the components involved and the transformations applied throughout the system.
Auditability of the full spectrum of behaviours introduced by {component interactions and their effects}
is thus hindered, whether components include AI functionality or not. %
The limited visibility across supply chains hinders various stakeholders (including developers, deployers, users, researchers,  policymakers, affected communities, and so on) who seek to understand, scrutinise, audit, govern, and regulate the systems, as part of a growing demand for AI accountability \cite{cobbe_understanding_2023, 10.1145/3531146.3533150, 10.1145/3442188.3445935}, transparency \cite{10.1145/3630106.3658985, Felzmann2020-rk, 10.1145/3531146.3533231}, and safety \cite{amodei2016concreteproblemsaisafety, Falco2021-mb, NEURIPS2024_7ebcdd0d}. Some work stresses the importance of being able to collect and trace data \cite{kroll_facct2021_traceability, singh_decisionprovenance}, not only as systems operate but also  as part of the development process \cite{cobbe_reviewability}, to assist audits and system interrogations.

One might argue that model `openness' could resolve some of these visibility challenges, as open-weight models offer greater access to model internals \cite{kim2024sealsuiteevaluatingapiuse}. Such openness can assist some system audits \cite{casper_black-box_2024, kembery2024positionpapermodelaccess}, as it enables auditors to examine model weights, analyse training processes, and access detailed behavioural indicators (e.g., log probabilities). However, even with an open FM, the surrounding scaffolding and infrastructure might operate as a black box, and individual stakeholders can still remain blind as to how various other parts of the systems operate and how other stakeholders have worked to shape implementation and behaviour.

The core `chained' transparency problem persists regardless as the level to which transparency is meaningful depends on the level of abstraction. While it is arguable that the highest abstraction level---examining the application as a whole%
---bypasses the chaining problem, it does not gain explanatory power, as outcomes are observable but one loses the understanding of how they come about.
When closely working with an open FM, open weights enable examining model internals,  but the API wrappers built on open models, intermediary transformations can remain opaque. As such, meaningful transparency varies by stakeholder position, persepective and audit objectives.

\textbf{Dynamism.} %
Algorithmic supply chains can entail components, configurations, and interactions that occur dynamically \cite{cobbe_understanding_2023, Chung2018DynamicSC, AtiehAli2024TheRB}. For example, the result of a content moderation API (AIaaS) may trigger different downstream processing paths: a positive result (community-guideline violation) then involves various sanction-related services, while a negative result means the content flows on for wide-spread distribution (see \autoref{fig:cascade_overall}).
This dynamism can manifest in different ways, at design time and at run time %
as components may activate conditionally based on particular inputs, configurations may adjust based on user behaviour or policy changes, or instantiations of the model, infrastructural aspects (e.g. host migrations), etc.

The move towards more `agentic AI' will likely exacerbate the dynamism of AI supply chains. For example, it may be that %
AI agents make runtime decisions about which tools and functions to invoke for specific tasks, instructions or inputs~\cite{wang2024gtabenchmarkgeneraltool, jones_automatically_2023, chen_towards_2024, tang_toolalpaca_2023}. %
Agentic AI will likely push supply chains %
that previously had more predetermined or deterministic pathways %
towards becoming highly dynamic systems, as the chain itself is constructed in a more ad-hoc fashion, `on-the-fly', during each invoked instantiation. As an example, an agent might dynamically attempt to discover APIs that might suit a task and ultimately select one, rather than %
always use a pre-determined one.

From the above we see that combinations of components can potentially be non-modular, highly dynamic, opaque for stakeholders, and limited in visibility for governing or auditing stakeholders.

\vspace{-0.5em}
\section{The Cascade} \label{sec:cascade}

We describe the  \textbf{cascade} as the process through which 
components interact, through a series of inputs\slash outputs, creating effects on data processing, adaptation and mediation for others.  %
That is, the cascade captures the cumulative interactions between components \cite{singh_decisionprovenance}, and provides a lens for understanding and examining their effects.\footnote{Note this is distinct from technical cascade models in machine learning or software engineering (e.g., cascading classifiers \cite{alpaydin1998cascading} or waterfall models \cite{petersen2009waterfall})} %
Cascades must be accounted to achieve systematic evaluations of safety, security, and other critical behaviours, as components could amplify \cite{hopkins2025aisupplychainsemerging}, dampen, or nullify (un)intended effects of multiple components. Harms can arise through these effects, as stakeholder goals are not met or actively circumvented.

\begin{figure}[h]
    \centering
    \includegraphics[width=0.9\linewidth]{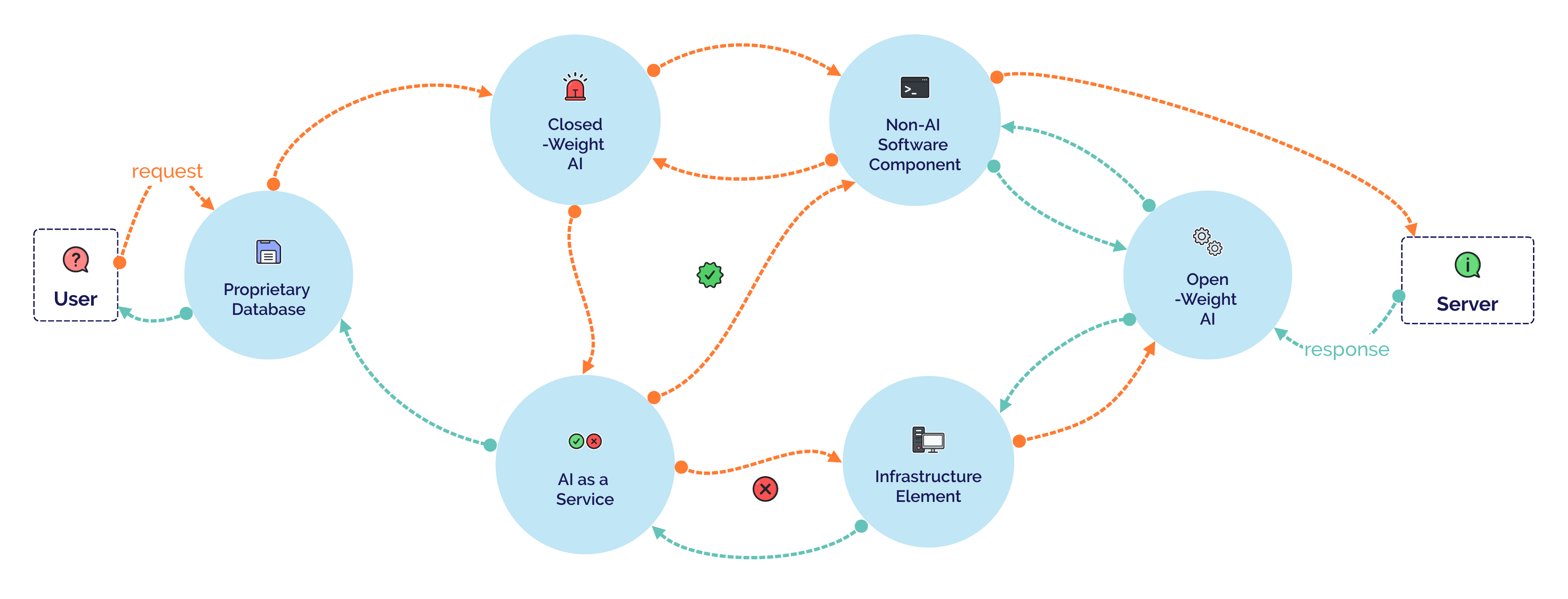}
    \caption{An overall view of possible implementations of cascades through AI supply chains. Orange (user request path) and green (response path) signal different component combinations. Illustrating the dynamic nature of the interactions, the AIaaS results (green and red symbol) directly effect what kind of component gets involved when processing a user request as the following step.}%
    \label{fig:cascade_overall}
    \vspace{-0.5em}
\end{figure}

To illustrate the cascade, we can consider a single data flow pathway through the supply chain (see \autoref{fig:cascade_overall}). %
Consider a corporate mental health support chatbot based on an LLM. An end-user writes, `I'm feeling overwhelmed' (orange path begins at user). A proprietary database retrieves their recent messages about project deadlines and late-night work sessions.
An AIaaS API analyzes whether this constitutes a mental health crisis and returns a \textit{positive }prediction. The crisis intervention plan is passed to the company server hosting the chatbot. To generate a response to the identified burnout case, the plan and relevant context are passed to an open-weight model (green path begins). This model generates a therapeutic response addressing the crisis, which is returned through a chain to the user. A simple expression of momentary stress escalates into a high-priority workplace mental health case with a formal report filed.

Different stakeholders may need to understand (and audit) \textit{different aspects} of the, again borrowing and now extending the metaphor\cite{cen2023ai}, \textit{soup-making process} depending on their positions in the supply chain -- not just the final dish. Whether components appear blended or remain distinct depends on context: which components are involved, to what degree, and at what level of abstraction they are examined. Some mechanisms may blend together and thus produce emergent effects, while others can remain separable and individually auditable.

Again, different stakeholders (and by extension audits) operate at \textit{different levels of abstraction}: patrons %
care about the final dish,  %
while chefs care about the cooking techniques, restaurant inspectors care about compliance with food safety regulations, and the person running the restaurant cares about making a profit off the soup. %
The cascade provides the analytical lens that supports these multi-level interactions \textit{based on each stakeholders' access, objectives, and responsibilities.}

To illustrate this analytical lens, we next describe two intersecting modalities: \textit{(i)} \textit{component} interactions affecting %
outcomes, while being \textit{(ii)} influenced by \textit{different stakeholders}.
These modalities are not exhaustive viewpoints for the cascade; rather they are useful for more precisely illustrating the intersecting effects, and showing their separate influences more clearly.

\vspace{-0.25em}
\subsection{Component Cascades}

As components in algorithmic supply chains interact, their effects can compound to shape system outputs in ways that can extend beyond the scope of each individual component (\S\ref{sec:definitions}). These effects can influence %
different objects: sometimes transforming how the model is configured or behaves, sometimes modifying data itself, and sometimes altering how information is presented or constrained.

\textbf{Transformation Types and Effect Composition.} We describe some forms of \textit{transformations} relevant in FM contexts%
, while acknowledging that comprehensive auditing would involve additional component effects %
based on infrastructure configurations, access controls, or logs~\cite{chappidi2025accountabilitycapturerecordkeepingsupport} (\S\ref{sec:definitions}):

\footnotesize
\begin{inlinenums}
    \small
    \item \textbf{Augmentative transformations} add information or capabilities (e.g., RAG systems \cite{sridhar2025regentretrievalaugmentedgeneralistagent}, tool use frameworks and abilities \cite{10.1145/3704435}, domain-specific information retrieval \cite{emde2025shhdontsaythat}), \newline
    \item \textbf{Behavioural transformations} modify how the model processes information (e.g., system prompts \cite{li_evaluating_2024, positionispower}, RLHF \cite{liu2024rlgptintegratingreinforcementlearning}, fine-tuning \cite{choi2024safetyaware}, reasoning \cite{asgari2024mmlupro, xue2024decompose}, Mixture-of-Experts paradigms \cite{cai_survey_2024}), \newline
    \item \textbf{End-user contextual transformations} (including personalisation) incorporate individual user information (e.g., conversation history \cite{noauthor_memory_2024}, personal preferences \cite{li2024personalizedlanguagemodelingpersonalized, zhao2025llmsrecognizepreferencesevaluating}, user-specific settings), \newline
    \item \textbf{Restrictive transformations} remove or suppress certain aspects of inputs or outputs (e.g., content moderation \cite{mahomed_auditing_2024}, safety filters \cite{dong_safeguarding_2024}, refusal mechanisms \cite{cao2024learnrefusemakinglarge}), \newline
    \item \textbf{Presentational transformations} alter the format, organization, or presentation of information (e.g., output formatting rules, response structure templates \cite{claudeSystemPrompts, openai_structuredoutputs}).
    \normalsize
\end{inlinenums}
\normalsize

These transformations can layer across the supply chain: FM developers set base constraints, deployers add business-specific modifications, and end-users apply personal preferences. %
Through the functionality that a component brings (i.e. the transformations they perform), and their interactions with other components, 
these transformations can reinforce, conflict with, or alter each other. As transformations accumulate, their combined effects become difficult to predict or to trace back to specific components.

\textbf{Implications for Auditing.} %
When audits consider components in isolation, they overlook how effects combine across the supply chain.
Audits must also examine both individual components and their effects in combination, so as to better reflect how system behaviour emerges.

\vspace{-0.25em}
\subsection{Stakeholder Cascade} \label{sec:cascadestakeholder}

We have described how different stakeholders (developers, deployers, end-users) %
will manage different components, which in a FM context might include: reasoning processes \cite{deepseek-ai_deepseek-r1_2025, wei_chain--thought_2023, noauthor_openai_nodate-1, NEURIPS2024_e4be7e98, chen2024unlockingcapabilitiesthoughtreasoning, zhang2024multimodalselfinstructsyntheticabstract}, memory functions \cite{noauthor_memory_2024}, knowledge bases \cite{fan_survey_2024, zhu2024knowagent, zhang_inftybench_2024}, tool use capabilities \cite{grattafiori_llama_2024, wang_what_2024, 10.1145/3704435}, and safety guideline adaptations \cite{mahomed_auditing_2024, dong2024safeguardinglargelanguagemodels, ngong2025protectingusersthemselvessafeguarding}. 

As different stakeholders control different socio-technical components, and therefore, different parts of the AI supply chain, %
the cascade can also be analysed through the viewpoint of stakeholders involved across supply chains. This is particularly relevant for identifying the party or parties that might be responsible for contributing to a particular system occurrence, and thus those who may need to respond (e.g. to repair, provide redress and restitution, etc). We discuss two characteristics: (i) \textit{visibility and control gaps} across AI supply chains that can undermine accountability and responsibility, and (ii)  \textit{emergent complexities}, that can hinder diagnosing harms and risks.

\textbf{Visibility and Control Gaps.}
Visibility is distributed over the components in an algorithmic supply chain. As each stakeholder (typically) only sees their portion (i.e. what they are able to control and manage)~\cite{Kroll_2021, singh_decisionprovenance, positionispower}, typically no entity is able to observe all components and their interactions %
\cite{nissenbaum_accountability_1996, cobbe_aiaas_2021}. 
Similarly, each entity can only control the components they are able to manage, meaning that control is distributed across the  AI supply chain~\cite{doi:10.1177/20539517231177620, cobbe_understanding_2023}. 
This can be illustrated by the `chain of command' \cite{openaiOpenAIModelspec} component: user prompts get processed through levels of authority, i.e. the %
{user} prompt is appended automatically by stored user preferences and can then be influenced by developer instructions or platform rules. %
Tracing which prompt change contributed to a final behaviour becomes challenging. Other interaction points and their effects are similarly difficult to detect and trace \cite{cobbe_aiaas_2021, cobbe_understanding_2023, brintrup_trustworthy_2023}.

Importantly, harms can emerge from cascade-driven effects that are not broadly visible or controllable. %
For example, safety measures might be weakened through other components: conflicting modifications (`X is good practice' vs `Don't do X!') could make the model less stable, or modifications dependent on multiple parties might fail. This could result in different harms: unsafe model outputs, discriminatory decisions, or security vulnerabilities. {These} harms are not only hard to measure and challenging to troubleshoot, they make accountability (more) ambiguous in terms of who may have contributed to what outcome. Ultimately, the nature of a cascade means that risks can emerge pertinent for the overall safety\slash security of a system.

\textbf{Emergent Complexity.}
When multiple components interact in ways no single actor can observe, unintended failure modes become inevitable, as systems grows too complex to maintain comprehensive oversight~\cite{perrow2011normal}. This brings complexity risks, i.e. failures that can emerge in any complex system, where interacting failures can occur despite efforts to prevent them \cite{dobbe_facct2022_systemsafetyAI}.

These and further effects can arise from combining %
a number of 
components, but particular behaviours may not appear when each component is audited in isolation. The high degree of dynamism in AI supply chains will likely increase such complexities, particularly agentic functions that may entail even more dynamic operations.
This matters for stakeholders as they may not be able to \textit{(i)} predict how their modifications will interact with others, nor \textit{(ii)} diagnose 
complexity risks, as tracing behaviour back to specific components\slash stakeholders becomes challenging (see the `accountability horizon'~\cite{cobbe_understanding_2023}).

\textbf{Implications for Auditing.} The combination of the limited visibility over the existence of components, their interactions, and their effects, together %
with 
the growing complexity in the AI ecosystem,  calls for auditing methods that can capture both individual components and their relationships 
within and across AI systems \cite{singh_decisionprovenance}. %
Such approaches should span stakeholder boundaries, enabling the systematic evaluation of components and their effects across the supply chain, and  enable  stakeholders to understand how their own %
contributions relate to, and are shaped by, those of others. %

\section{The Current Foundation Model AI Auditing Landscape}
\label{sec:auditlandscape}

Having outlined auditing requirements from a general cascade perspective, we now examine the degree to which current LLM-oriented audits engage with these implications. %
We focus on LLMs for several reasons: they are deployed across diverse domains, and are conceptualized as general-purpose FMs with wide-ranging supply chains. As such, they are a dominant focus of research on auditing, where this degree of maturity helps support our analysis and ways forward. 
We identify that the current landscape of audits of LLM systems centers on two main areas: \textit{(i)} capabilities of FMs, including audits of particular components, and \textit{(ii)} domain-adapted applications.
This brief survey of the literature on LLM audits,  grounds our {analysis} %
in existing approaches and identify where they align with, or fall short of, the implications of cascades.

\subsection{Foundation Model Audits} \label{sec:auditfm}

Foundation models raise significant societal, ethical, and accountability concerns across their development and deployment \cite{dominguez_hernandez_mapping_2024, diberardino_algorithmic_2024, wu_sustainable_2022, weidinger_ethical_2021, chan_harms_2023, gornet_mapping_2024, bogiatzis-gibbons_beyond_2024}. These concerns motivate audits at multiple stages and by diverse stakeholders, e.g., developers test models before release \cite{grattafiori_llama_2024, noauthor_introducing_2024}, and third-party researchers evaluate publicly available models \cite{AISafetyInstitute_evals}.

\textbf{Capability Audits.} FM audits predominantly focus on evaluating model capabilities in controlled, reproducible settings through standardised benchmarks. These audits assess fundamental abilities like question-answering \cite{touvron_llama_2023, grattafiori_llama_2024, jiang_mistral_2023, 10.1145/3663363, noauthor_claude_nodate}, robustness to adversarial attacks \cite{chouldechova2024red, fan2024resilientefficientllmscomparative}, and performance on specific tasks \cite{liang_holistic_2023, noauthor_introducing_2024, hendrycks_measuring_2021, noauthor_openai_nodate-1, noauthor_openai_nodate-2}. Benchmarks therefore enable systematic comparison across models by defining specific ranges of tasks probing particular behaviours \cite{qin_sysbench_2024, bai_benchmarking_2023, souly_strongreject_2024, chollet_measure_2019, huang2025effibenchbenchmarkingefficiencyautomatically, NEURIPS2024_085185ea, NEURIPS2024_1bdcb065, liang2023holisticevaluationlanguagemodels, luo2025bigbenchunifiedbenchmarkevaluating}. One prominent example is MMLU \cite{hendrycks2021measuringmassivemultitasklanguage}, which aims at measuring the broad ability to assess `language understanding' in a range of different tasks.  

Beyond broad capability assessments, evaluations target specific properties of the FM: factual accuracy \cite{krishna2024genaudit}, bias \cite{parrish_bbq_2022, tamkin_evaluating_2023, gupta_bias_2024, salinas_whats_2025, rottger_xstest_2024}, privacy compliance \cite{panda_privacy_2024}, reasoning abilities \cite{chollet_measure_2019, deepseek-ai_deepseek-r1_2025, wei_chain--thought_2023, noauthor_introducing_2024, NEURIPS2024_b631da75, NEURIPS2024_e5d1eaad}, knowledge representation \cite{petroni2019language, safavi2021relational, hernandez2023measuring}, instruction-following \cite{zhou_instruction-following_2023, li_evaluating_2024}, and `alignment' to specified values or value systems \cite{mazeika_utility_2025, greenblatt_alignment_2024, ji2024languagemodelsresistalignment}.

\textbf{Component-Specific Audits.} \label{sec:auditcomponent}
FM audits may \textit{also} include examinations of certain technical components, e.g., model weights, architectures, and training data, but access remains limited for particular platform elements. Current research mostly examines components in isolation, focusing on one component, mostly leaving others unchanged to isolate effects.
Examples include evaluations of system prompts \cite{mu_closer_2024, qin_sysbench_2024, mukherjee_orca_2023, jiang_personallm_2024, jiang_mistral_2023, positionispower}, safety guardrails \cite{dong2024safeguardinglargelanguagemodels, ren2025stepbystepmasteryenhancingsoft, mahomed_auditing_2024, chouldechova2024red, rengarajan2024imitation, souly_strongreject_2024, wei_jailbreak_2024, chao_jailbreaking_2024}, or more technically focused audits, i.e., different quantization and compression techniques \cite{lin2024duquant, kirsten2024the, ji2024languagemodelsresistalignment, malinovskii2024pvtuningstraightthroughestimationextreme}, or RLHF methods \cite{ghosh2024aegis, liang2024rlhs, amortila2024reinforcement, barnhart2025aligningwhatlimitsrlhf, barnhart-etal-2025-aligning, choi2024safetyaware}.

\textbf{Broader FM Implications.} 
As mentioned above, research also examines the implications of FMs on the AI ecosystem, be it in development~\cite{mitchell2025fullyautonomousaiagents, konidena_ethical_2024, orr_ai_2024} or deployment~\cite{Putri_Tran_2023, MLSYS2022_462211f6, luccioni_power_2024, eyuboglu_model_2024, 10.1145/3544548.3581555} contexts.
This includes studies on environmental costs, labour practices, and the (possible) impact of different accountability or `responsible AI' practices ~\cite{10.1145/3442188.3445922, weidinger_ethical_2021, brundage_toward_2020, liu2024trustworthyllmssurveyguideline, 10.1145/3644815.3644959}.%
Further, some analyses examine the broader implications of FMs for particular sectors, such as medical applications \cite{mclennan2022embedded} or law enforcement \cite{berk2021artificial}, without looking at specific implemented systems.

\subsection{Application Audits} \label{sec:auditdomain}

Audits at the application-level evaluate systems in their entirety, and or deployed systems \textit{within real-world contexts} \cite{Cobbe2024-xx}. They typically focus on specific user experiences \cite{field_examining_2023, gadiraju_i_2023, mayworm_misgendered_2024, cheong2024not}, or broader interaction patterns \cite{widder_its_2023, madaio_learning_2024}, with methods like user studies \cite{10.1145/3613905.3650732, 10.1145/3706598.3713218}, or UX-based evaluations \cite{10.1145/3706599.3716220, 10.1145/3706598.3713200, 10.1145/3706598.3713140}. Assessments of this kind can generate targeted design recommendations \cite{10.1145/3173574.3174014, sterz_quest_2024} that address particular application requirements and user needs \cite{10.1145/3687025, hancock_trouble_2024, ball-burack_differential_2021, rivera_escalation_2024}. Further, these audits can focus on domain-specific performance metrics \cite{schor_meaningful_2024, qiu_llm-based_2024, li2024legalagentbench, moayeri2024worldbench, bertsimas2024m3hmultimodalmultitaskmachine, kim2024mdagents} through such methods as simulated task completions \cite{doi:10.1142/S1793962324300024, valizadeh-parde-2022-ai, 10.1007/978-3-030-52240-7_13}, or compliance audits~\cite{PANIGUTTI2021102657}.%

While these approaches provide valuable insights within their domains, they typically concentrate on final outputs and user interactions, while treating the FM (or other aspects of the system) as a black box~\cite{casper_black-box_2024} or as one component instead of multiple connected components~\cite{cobbe_understanding_2023, cobbe_aiaas_2021, brintrup2023trustworthyresponsibleethicalai}.
Of course, developers also routinely test applications for security vulnerabilities, and safety failures in pre-deployment tests. However, these typically occur in envisaged usage contexts and may not always consider interactions that only emerge when systems are deployed as part of a broader algorithmic supply chain, particularly interactions with other systems that developers may not be aware of\slash control.

\section{Auditing Gaps} \label{sec:auditgaps}

Current FM audit approaches would struggle to capture all cascading effects. First, FM audits focus mostly on model abilities, or examine components like system prompts or tool use alongside or in contrast to base models. Yet, they typically assess these components separately\slash disconnected from their specific deployments. %

At the same time, application audits focus on final system behaviours in deployed contexts. However, such audits typically %
observe or capture only some of the components %
throughout the supply chain;
while they may register some cascade effects, they often cannot determine that these effects originate from a cascade, nor identify their specific causes, and may miss behaviours that are not apparent or envisaged at that level of testing.%
\footnote{
Again, what counts as an `application' versus a `component' can be perspective-dependent. A service or component could  be considered either an application (the object being audited) or a component (a part within a larger system) depending on the level of abstraction and stakeholder position (\S\ref{sec:definitions}). As \textit{one actor's component can be another actor's application}%
, auditors can zoom in and out as appropriate given surrounding factors.}
Limited auditing access to or knowledge of %
particular components means that other components---including those added by downstream deployers or ones activated conditionally through other runtime factors---are not observed. In this way %
certain specific audits might capture that something happened, but not necessarily \textit{why} or \textit{how} the cascade of components produced that outcome. %

Although %
existing auditing components capture many effects, they often do %
not directly consider how the technical components interact with {each other}, %
how they influence each other, and what they might cause. Therefore, \textbf{AI evaluations also need holistic and systematic evaluations taking cascades into account}, i.e., examining component interactions at different levels of abstraction and from different stakeholder viewpoints. %
As such, and returning to characteristics of AI supply chains (\S\ref{sec:aisupplychains}), %
auditing regimes would need to account for:

\noindent\textbf{Opacity.} As the visibility across stakeholders is limited, the complex outcomes created by components amplifying, dampening, or nullifying each others effects
dynamically throughout AI supply chains remain under-explored. %
Current auditing approaches will often miss effects, with real-world implications, brought about by deployments that involve a range of interacting components.

\noindent\textbf{Non-modularity.} AI supply chains exhibit non-modular interactions that current audits struggle to capture. As components can form feedback loops, activate conditionally, compound through repeated calls, or show emergent behaviours (perhaps only) when combined with specific other components, audits that examine components separately \textit{or} the whole application without accounting for component interactions %
can miss these effects. Comprehensive audits must therefore (help to) trace how components influence each other and how effects emerge across complex pathways. %

\noindent\textbf{Dynamism.}
We discussed how the interactions between components can be dynamic, in that different components can come together in different ways to produce certain behaviours, perhaps even when part of the same application.
As such, audits will need to consider %
the 
range of potential interactions that components might have. Again, systems are becoming more dynamic with agentic AI, where agents can have higher propensity and more degrees of freedom to create new supply chains dynamically at runtime, e.g. by discovering and selecting different components and sequences for each particular task. This makes audits that account for dynamism more challenging.  %
Moreover, when multiple agents operate together, their choices can influence each other's subsequent decisions, making audits of single pathways insufficient.

\section{Approaches to Cascade Auditing} \label{sec:cascaud}

We now propose some research directions across the two cascade dimensions (component and stakeholder), %
 and %
 highlight some arising tensions and opportunities for research.

\subsection{Technical Research Directions} \label{sec:cascaudtech}

The dynamics of cascades pose several technical challenges for auditing and motivate a set of technical research directions, which we describe in \autoref{tab:technicalimplications}: 

\begin{table}[h]
    \centering
    \footnotesize
    \caption{Critical Technical Implications of %
    Component Cascades}
    \begin{tabular}{p{0.2\linewidth}p{0.75\linewidth}}
    \textbf{Semantic Altering} & Each component can alter an input's meaning, intent, and structure beyond surface-level changes, potentially distorting user intent. \\
    \hline
    \textbf{Authority Hierarchy} & Components operate in a hierarchy of authority, where higher-level transformations can override or nullify lower-level changes, obscuring accountability and making it difficult to trace the source of harmful outcomes. \\
    \hline
    \textbf{Invisible Processing} & These transformations occur invisibly to most stakeholders, preventing effective oversight of compounding changes throughout the system. \\
    \hline
    \textbf{Unexpected \newline Behaviours}%
    & Component interactions can create unexpected behaviours that escape both FM and application-level audits but significantly impact system responses. \\
    \hline
    \textbf{Scaling Resistance} & Research shows that FMs become less willing to modify their core behaviours and preferences as they scale up \cite{mazeika_utility_2025}, suggesting fundamental limits to how much intermediate components can or cannot reshape model responses. \\
    \end{tabular}
    \label{tab:technicalimplications}
\end{table}

The above highlights why the technical cascade requires evaluation approaches that can 
\textit{(i)}
track or otherwise give insights regarding components and %
\textit{(ii)}
measure the cumulative effects that result and flow from components and their interactions with others. %

New and adapted audit and evaluation methodologies could help address some of these technical considerations; for example, methods that enable lineage tracking or tracing behavioural changes from model release to end-user interaction \cite{singh_decisionprovenance, palla_policy-as-prompt_2025, Kroll_2021}, %
which can help capture crucial transformations that could alter LLM behaviour in safety, security, and harm-critical ways \cite{fortuneAIpoweredCoding, bbcParentsTeenager}.

Against this backdrop, we identify some related %
research directions that are promising for the current (and future) AI ecosystem:

\footnotesize
\textbullet\ \textbf{Methods to track data flow} and transformations specifically through component interaction points, \newline
\textbullet\ \textbf{Methods that scale across supply chains} as the growing sophistication of LLM supply chains will result in introducing additional components, interaction points, and cumulative effects, \newline
\textbullet\ \textbf{Fairness, reliability, and accountability implications} of component hierarchies, \newline
\textbullet\ \textbf{Evaluation of system changes} through varying configurations, %
\newline
\textbullet\ \textbf{Assessment of safety component robustness} across interaction points.
\normalsize

Additionally, some promising research indicating %
possible ways forward exists, including traceability \cite{Kroll_2021}, reviewability \cite{cobbe_reviewability}, decision provenance \cite{singh_decisionprovenance}, or methods from systems theory \cite{dobbe_facct2022_systemsafetyAI, leveson_2004_systemstheorysafety}. %

\subsection{Human-Centered and Interdisciplinary Approaches} \label{sec:cascaudhuman}

Human-centered examinations of how stakeholders configure, interpret, and interact across AI supply chains and cascades are also necessary. There is a role for research to %
examine %
how stakeholders (developers\slash deployers, regulators\slash oversight bodies, users, and affected communities) are influenced by various components, how various stakeholders can better understand cascades, and how the stakeholders themselves shape the {everyday operations of deployed systems}.

\textit{Interdisciplinary approaches}, bridging technical and social dimensions, will be key in helping to 
examine how stakeholders engage with the components they manage (and indeed, those they do not manage). 
Some promising research directions include: 

\footnotesize
\textbullet\ \textbf{Stakeholder-focused studies} investigating how various stakeholders interpret, configure, and respond to observable or hidden interaction points from their specific viewpoint (see ~\cite{balayn2025unpacking, balayn2024understandingstakeholdersperceptionsneeds}), \newline
\textbullet\ \textbf{Participatory audit frameworks} that combine stakeholder expertise with technical analysis to evaluate component interactions from multiple perspectives, \newline
\textbullet\ \textbf{Mixed-method investigations} tracking technical transformations and their impact on different user groups, \newline
\textbullet\ \textbf{Derivation of responsible use guidelines} that help stakeholders understand how their component configurations affect others in the supply chain and implement more safe and robust AI-based systems.
\normalsize

\subsection{Tensions and Resulting Research Opportunities} \label{sec:tensions}

We highlight several key tensions that, in turn, also represent opportunities for research:

\textbf{System Tailoring.} 
Components operate to provide functionality and to tailor outputs for specific contexts. In this way, components can both enhance functionality, as well as potentially raise risks. Auditing that accounts cascades allows for a better understanding of, and therefore the ability to respond to issues that particular components or their combinations might bring. After building such understandings, careful consideration regarding the interventions resulting from audits is needed, as interventions may need to balance functional benefits and potential risks of specific components, their behaviours, and their interactions.

\textbf{Balanced Transparency Needs.} Increased transparency through component auditing, while helping in managing risks, can also raise potential risks of their own. This can include security and safety risks, e.g. by facilitating adversarial attacks, such as where revealing safety-related system prompts could enable tailor-made jailbreak attacks. Selective transparency measures and controlled disclosures can potentially help enable a balancing of oversight and information sensitivity~\cite{norval_discbydesign}.

\textbf{Inherent System Complexity.} %
Although similar coordination challenges and emergent outcomes are present across various complex systems, AI supply chains raise additional complexities given their non-modular and dynamic nature. As AI supply chains are very opaque in practice, from model weights to training processes, this should be considered in real-world auditing challenges.

\section{Conclusion} \label{sec:conclusion}

Socio-technical components drive AI supply chains, shaping system outcomes and behaviours through complex interactions that neither FM- nor application-level audits can fully capture. 

We have described a %
\textit{cascade} that creates systematic evaluation gaps: the %
component modality, where effects compound through component interactions, and the stakeholder modality, where distributed control and visibility across multiple stakeholders means no single entity has the complete knowledge or control.
{Through introducing the cascade as an object of analysis}, we outline 
technical-, stakeholder-, and interdisciplinary-oriented approaches for examining components, their interactions, and downstream outcomes. While our analysis focuses on LLM-based systems, the concepts should apply to other AI systems that involve a range of interacting components.  %

In all, the cascade offers a lens that highlights %
key auditing gaps, raising important and timely challenges for research and practice, as AI systems---particularly those general-purpose---will increasingly impact everyday life.

\section{Statement of Contribution} \label{sec:statementcontribution}

This paper identifies a critical gap in AI safety and security evaluations: current approaches focus on FMs or deployed applications while missing components and their interactions. 
We introduce the \textit{cascade} as an object of analysis. %
In doing so, we provide the foundation for further research directions examining interacting behaviours that current auditing approaches miss.

\clearpage
\bibliographystyle{unsrtnat}
\bibliography{bib/heal_2025, bib/heal_2025_extra}

\clearpage
\appendix

\end{document}